\documentclass[twocolumn,english,aps,prd,amsmath,amssymb,reprint,floatfix,nofootinbib,showkeys]{revtex4-2}

\usepackage{graphicx}
\usepackage{dcolumn}
\usepackage{bm}
\usepackage[colorlinks = true,citecolor=blue,linkcolor = blue,urlcolor=blue]{hyperref}

\usepackage[noabbrev]{cleveref}
\usepackage{graphicx}	
\usepackage{amsmath}	
\usepackage{amssymb}	
\usepackage{wasysym}    
\usepackage{mathtools}  
\usepackage{aas_macros}
\usepackage{array,multirow,graphicx}  

\newcommand{\modelz}{M_{\rm 0}}
\newcommand{\modelo}{M_{\rm 1}}

\newcommand{\post}{\mathcal{P}}
\newcommand{\prior}{\pi}
\newcommand{\like}{\mathcal{L}}
\newcommand{\Z}{\mathcal{Z}}

\newcommand{\br}{\mathcal{K}}

\newcommand{\enf}{f_{\rm EN}(D)}

\begin{document}


\title{Fully Bayesian Forecasts with Evidence Networks}

\author{T. Gessey-Jones}
    \email{tg400@cam.ac.uk}
    \affiliation{Astrophysics Group, Cavendish Laboratory, J.J. Thomson Avenue, Cambridge, CB3 0HE, UK}
    \affiliation{Kavli Institute for Cosmology, Madingley Road, Cambridge, CB3 0HA, UK}
    
\author{W. J. Handley}
    \email{wh260@mrao.cam.ac.uk}
    \affiliation{Astrophysics Group, Cavendish Laboratory, J.J. Thomson Avenue, Cambridge, CB3 0HE, UK}
    \affiliation{Kavli Institute for Cosmology, Madingley Road, Cambridge, CB3 0HA, UK}

\date{\today}

\begin{abstract}
Sensitivity forecasts inform the design of experiments and the direction of theoretical efforts. 
To arrive at representative results, Bayesian forecasts should marginalize their conclusions over uncertain parameters and noise realizations rather than picking fiducial values. 
However, this is typically computationally infeasible with current methods for forecasts of an experiment's ability to distinguish between competing models. 
We thus propose a novel simulation-based methodology capable of providing expedient and rigorous Bayesian model comparison forecasts without relying on restrictive assumptions.
\end{abstract}

\maketitle

\textit{Introduction.}---
\citet{EN} recently proposed Evidence Networks.
These are classifier artificial neural networks trained on data generated from two competing models using specific loss functions, so that the output of the network coverges to an invertible function of the Bayes ratio~\citep{Mackay_2003} between the two models. 
Consequently, these networks can be used for simulation-based Bayesian model comparison as illustrated in various examples by those authors. 
Here, we highlight a promising use case of these networks not mentioned in that original paper, forecasts of Bayesian model comparison analyses from upcoming experiments.

Forecasts are an essential part of science.
Estimates of required sensitivities guide the design of experiments, and whether these experiments can probe different theoretical models informs which models time and resources are spent developing. 
Furthermore, funding agencies use anticipated scientific conclusions as part of their decision-making. 
As a result, fast, accurate, and reliable forecasting techniques play a vital role in the modern scientific method.

Due to its importance, many techniques exist to perform forecasts~\citep[e.g.][]{Fisher_1922, Trotta_2007b, Sellentin_2014, Alvey_2022a, Alvey_2022b, Ryan_2023}. 
Within a Bayesian paradigm, arguably the most accurate forecasting technique is to generate mock data and perform Bayesian analysis on the data as if it were experimental data~\citep[e.g.][]{Anstey_2021, Rieck_2023}.
However, such an analysis is typically computationally costly, limiting its application to a few mock data realizations.
Hence, these analyses often break a central tenant of Bayesian statistics by drawing conclusions based on a small number of fiducial parameter values, which may lead to erroneous conclusions if the mock data used is not representative of reality.

To avoid this issue, the conclusions of Bayesian forecasts should be marginalized over uncertain parameters and noise realizations to perform a fully Bayesian forecast~\citep{Mukherjee_2006, Sivia_2006, Pahud_2006, Trotta_2007, Trotta_2007b, Gelman_2014, Leung_2017}.
However, performing such an analysis in practice is often infeasible with the tools used to analyse experimental data sets, e.g., MCMC and nested sampling, due to the aforementioned computational cost.
Hence, Fisher forecasts~\citep{Fisher_1922}, for parameter constraint estimates, or Savage–Dickey forecasts~\citep{Dickey_1971, Trotta_2007b}, for model comparison projections, are commonly employed across astronomy and astrophysics~\citep{Tegmark_1997, DETF, Trotta_2007b, 2007_Seo, Vallisneri_2008, More_2013, DiDio_2014, Zhai_2017, Bonvin_2018, Simons, Euclid_2020, dAssignies_2023, Mason_2023} to cover the potential data space and investigate how conclusions vary.
Unfortunately, the Gaussianity assumption these forecasts rely upon does not always hold, limiting their reliability~\citep[e.g.][]{Perotto_2006, Wolz_2012}.
Furthermore, Savage–Dickey forecasts can only be used for nested models (when one of the models is a special case of the other), limiting their applicability~\citep{Trotta_2007}.

Current analysis methodologies for Bayesian model comparison are thus either too slow for fully Bayesian forecasts to be feasible e.g., nested sampling, or only applicable in specific cases, e.g., Savage–Dickey forecasts.
In this letter, we propose utilizing simulation-based inference in the form of Evidence Networks~\citep{EN} to overcome the constraints of current methods~\citep{Mukherjee_2006, Trotta_2007b}, enabling expedient, fully Bayesian forecasts on any scientific question formulated as a condition on the Bayes ratio between two models.
Such questions include whether a signal can be detected from within noise, whether two competing theories can be distinguished by expected data, or at what sensitivity level will a piece of additional physics be required within a model?
We then demonstrate the technique by finding the \textit{a priori} chance of detection of the global 21-cm signal by a REACH-like experiment~\citep{REACH}, an analysis that would have been computationally impracticable using traditional methods.

\textit{Bayesian Forecasting.}---
A traditional Bayesian analysis~\citep[see][for a more detailed discussion]{Mackay_2003} starts with a model $M$ that takes some parameters $\theta$.
From the model and experimental considerations, a likelihood of observing data $D$ is constructed $\like(D | \theta, M)$.
In addition, an initial measure, the prior $\prior(\theta | M)$, is put over the parameter space to quantify the \textit{a priori} knowledge of the parameter values.
Then given some (mock) observed data, the measure over the parameter space is updated via an application of Bayes' theorem
\begin{equation}~\label{eqn:bayes}
    \post(\theta | D, M) = \frac{\like(D | \theta, M) \prior(\theta | M)}{\Z},
\end{equation}
into the \textit{a posteriori} knowledge of the parameter values $\post(\theta | D, M)$, called the posterior. 
Here $\Z$ is the Bayesian evidence
\begin{equation}
    \Z = \int \like(D | \theta, M) \prior(\theta | M) d\theta = P(D | M),
\end{equation}
which serves both as a normalization constant and as a natural goodness-of-fit statistic for comparing competing models of the same data.
Now suppose there are competing models, $\modelz$ and $\modelo$, with corresponding Bayesian evidences $\Z_{\rm i}$.
If we have an \textit{a priori} belief in each model of $P(M_{\rm i})$ then for each model the \textit{a posteriori} belief  $P(M_{\rm i}| D)$ in the model is found using a further application of Bayes' theorem
\begin{equation}~\label{eqn:bayes_on_models}
    P(M_{\rm i}| D) = \frac{P(D | M_{\rm i}) P(M_{\rm i})}{P(D)} = \frac{\Z_{\rm i} P(M_{\rm i})}{P(D)}.
\end{equation}
Cancelling $P(D)$ between the above with $i = 0$ and $i = 1$ then gives the Bayesian Model Comparison equation
\begin{equation}~\label{eqn:bayes_model_prob_update}
    \frac{P(\modelo| D)}{P(\modelz | D)} = \frac{\Z_{\rm 1} P(\modelo)}{\Z_{\rm 0} P(\modelz)}.
\end{equation}
This equation shows that after observing some data $D$ the relative belief in the two models is updated with the ratio of their evidences, e.g.\ the belief in the higher evidence model increases, and the belief in the lower evidence model belief decreases. 
It is common to assume initially that the two models are equally likely $P(\modelo) = P(\modelz)$ in which case \cref{eqn:bayes_model_prob_update} simplifies to
\begin{equation}~\label{eqn:bayes_ratio}
    \br = \frac{P(\modelo| D)}{P(\modelz | D)} = \frac{\Z_{\rm 1} }{\Z_{\rm 0}},
\end{equation}
so that the Bayes ratio $\br$ of the model evidences is simply the relative belief in model $1$ compared to model $0$ after observing some data. 
The above can also be rearranged to find the posterior belief in a model 
\begin{equation}~\label{eqn:model_posterior_absolute}
    P(\modelo| D) = \frac{\br}{1 + \br},
\end{equation}
which can be compared to a statistical threshold to determine if one model is significantly preferred over the other.

With the mathematical notation of Bayesian analysis established, we can return to the central topic of this letter: forecasts.
For many scientific applications~\citep{Mukherjee_2006, Trotta_2007}, the question of interest is best posed as whether a proposed experiment could distinguish between two competing models.
A traditional Bayesian forecast of such a question would go through the following steps. 
First, the models and corresponding likelihoods are established.
Then, mock data is simulated from one of the models by choosing some fiducial parameter values $\Tilde{\theta}$, adding noise $\eta$, and experimental effects.
$\Z$ is evaluated for each model treating the mock data as you would actual observations from which $\br$ (or equivalently $P(M_{\rm i}| D)$) is calculated. 
Finally, $\br$ is compared to some condition to determine if the favouring of a model is statistically significant and so the models are distinguishable.

The conclusion of the outlined forecast is hence based on a condition on $\br$ of the form $\br > \br_{\rm crit}$ with $\br_{\rm crit}$ a statistical significance threshold. 
However, $\br$ is conditional on the fiducial parameters chosen to generate the mock data and the noise added, so it is more accurate to state
\begin{equation}~\label{eqn:partially_bayesian_forecast}
    \br\left(\Tilde{\theta}, \eta\right) > \br_{\rm crit}.
\end{equation}
Using such a criterion presents two major issues (previously identified and discussed in \citet{Mukherjee_2006} and \citet{Trotta_2007b}). 
Firstly, the result is dependent on random noise $\eta$, leading to differences in conclusion purely based on chance.
Secondly, these conclusions stem from a single point in the parameter space $\Tilde{\theta}$, leading to differences based on the choice of fiducial model (anathema to Bayesian statistics).
A central tenant of Bayesian statistics being that particular sets of parameter values, even those that maximize the likelihood or posterior, have zero measure and thus are vanishingly unlikely and should not be used to draw conclusions.
Instead, in Bayesian statistics, conclusions should be reached by marginalizing over the parameter space weighted by some measure, typically the prior or posterior. 
This issue is further compounded in the case of forecasts as there is often limited advance knowledge of the parameter values leading to the conclusion drawn potentially varying widely between choices of $\Tilde{\theta}$.

A rigorously Bayesian approach would be to marginalize the condition in \cref{eqn:partially_bayesian_forecast} over $\Tilde{\theta}$ and $\eta$. 
Such a procedure would give the expected probability of the condition holding and thus of drawing some scientific conclusion rather than a binary yes or no answer.
Since the prior represents our knowledge of the parameters before any data is measured, it is the natural measure for a forecasting analysis.
So, for our example, we should calculate
\begin{equation}~\label{eqn:fully_bayesian_distinguish}
    \mathbb{E}(\textrm{Distinguish Models}) = \left\langle \br\left(\Tilde{\theta}, \eta\right) > \br_{\rm crit} \right\rangle_{\eta, \prior(\Tilde{\theta})},
\end{equation}
to find a fully Bayesian forecast of our expected chances of distinguishing between the models at the statistical significance set by $\br_{\rm crit}$.
More generally, the condition in \cref{eqn:fully_bayesian_distinguish} could be replaced with any other condition on $\br$ to draw a scientific conclusion
\begin{equation}~\label{eqn:fully_bayesian_condition}
    \mathbb{E}(\textrm{Drawing Conclusion}) = \left\langle \textrm{Condition}\left[\br(\Tilde{\theta}, \eta)\right] \right\rangle_{\eta, \prior(\Tilde{\theta})},
\end{equation}
to perform a fully Bayesian forecast on drawing said conclusion. 
In addition, \cref{eqn:fully_bayesian_condition} can be modified to produce fully Bayesian forecasts for a broader range of scientific questions. 
For example, when appropriate to the question being posed, the marginalization should be performed over the model prior as well (this is equivalent to marginalizing over the predictive posterior odds distribution introduced in \citet{Trotta_2007b}), or over conditional priors. 
Thus, fully Bayesian forecasts can in theory be used to give principled and interpretable answers to a range of forecasting questions.

However, performing the above average over the parameters and noise (and potentially models) in practice would require thousands or millions of mock data sets and, in turn, thousands or millions of Bayes ratio calculations.
\citet{Mukherjee_2006} proposed using Nested Sampling~\citep{Skilling_2004, NS_Primer} to calculate the $\br$ values for fully Bayesian model comparison forecasts, but in practice, this is typically
prohibitively computationally expensive, even for simple problems.
To circumvent these computational limitations \citet{Trotta_2007} proposed using the Savage–Dickey density ratio~\citep{Dickey_1971, Verdinelli_1995} to rapidly evaluate $\br$ values between nested models, facilitating a fully Bayesian forecast of whether the \textit{Planck} satellite could detect a deviation in the scalar spectral index of primordial perturbations $n_{\rm s}$ from 1.
The use of the Savage–Dickey density ratio for model comparison forecasts is analogous to the widespread usage of Fisher forecasts to perform parameter constraint forecasts over uncertain data spaces.
Both techniques utilize the analytic results available for linear models and Gaussian likelihoods to calculate approximate analytic parameter posteriors~\citep[e.g.][]{Tegmark_1997} or the Bayes ratio between models.
However, the reliability of these results is conditional on the accuracy of the implicit linearization and Gaussianity assumptions~\citep[see][]{Perotto_2006, Wolz_2012}.
Furthermore, usage of the Savage–Dickey density ratio requires the models being compared to be nested.
Thus, if we had nested models, an explicit likelihood, and knew the above assumptions to hold well, the Savage–Dickey density ratio would suffice to perform fully Bayesian model comparison forecasts.
However, these requirements significantly limit the usage of such a methodology. 
Reliable and widely applicable fully Bayesian forecasts will hence require a novel methodology that maintains the expedience of Savage–Dickey forecasts but is applicable to non-nested models and avoids the same assumptions.

\textit{Evidence Networks.}---
Evidence networks are a type of classifier artificial neural networks introduced recently in \citet{EN}.
They take in simulated data $D$ and output a single value $\enf$, with training performed on data generated from two models (labels $m = 1$ and $m = 0$).
The principal insight with these networks is that for a broad category of choices of loss function the evidence network's value converges toward an invertible function of the Bayes ratio $\br(D)$ between model $1$ and model $0$.
Hence, if such a network is converged, the Bayes ratio between the two models for any set of `observed' data can be directly calculated from the network's output.

As established in the previous section, efficient evaluation of $\br$ for a wide range of mock data is the main obstacle to performing practicable fully Bayesian forecasts. 
Since evaluating a neural network on a GPU is orders of magnitude faster than direct Bayesian evidence calculation techniques, utilizing evidence networks may circumvent this computational limitation. 
We thus propose an alternative evidence-network-based methodology for performing fully Bayesian forecasts:
\begin{itemize}
    \item First, create simulators of mock data from the two competing models (including experimental considerations such as noise or selection effects).
    \item Then, using the two simulators, generate a training set and validation set of labelled mock data and train the evidence network.
    \item Validate the evidence network using a blind coverage test~\citep{EN} to verify the network's accuracy and, if possible, also compare its output to a sample of $\br$ values calculated from traditional Bayesian techniques. This step is essential, as for data manifolds too complex to be classified by the chosen network architecture, or for insufficient training data sets, the converged network will not correspond to an accurate calculation of $\br$\footnote{In the example discussed in the next section we found a blind coverage test revealed the network was consistently underconfident (e.g., conservative) in these two scenarios.}. If validation indicates the network has not accurately converged, the network architecture or training will need to be refined and the previous steps repeated.
    \item Finally, evaluate \cref{eqn:fully_bayesian_condition}, or equivalent, using the network and previously developed simulators to evaluate $\br$ over $\prior$ and $\eta$ (and potentially $M_{\rm i}$) efficiently. 
\end{itemize}
We anticipate this approach to be highly performant relative to the traditional Bayesian approach since each $\br$ evaluation no longer requires an exploration of the parameter space.
Furthermore, our methodology does not require the models used to be nested or to satisfy the restrictive assumptions of the Savage–Dickey forecasting method; and only needs efficient mock data simulators rather than explicit likelihoods.
As a result, it can even be applied in simulation-based inference contexts where no closed-form likelihood exists.
We thus also anticipate it to be widely applicable.

\textit{An Application to Cosmology.}---
We have motivated fully Bayesian forecasts and argued that evidence networks may make performing one feasible.
Here, we demonstrate this feasibility while also illustrating some of the insights gained from such an analysis. 

The problem we tackle is determining the expected chances of a 21-cm global signal experiment making a definitive detection. 
Through the redshift evolution of the 21-cm global signal the thermal evolution of the Universe from recombination to reionization can be traced, giving insights into cosmology and structure formation~\citep[see][for a more detailed introduction to the field]{Furlanetto_2006}. 
However, measurement of the signal is challenging as its expected magnitude is five orders of magnitude below that of galactic foregrounds. 
As a result, there are currently no definitive detections of the global 21-cm signal.
The EDGES 2 experiment claimed a signal detection~\citep{EDGES}, but this is disputed~\citep[e.g.][]{Hills_2018} due to the signal not matching theoretical expectations~\citep[e.g.][]{Barkana_2018}, being better fit by the presence of a systematic~\citep[e.g.][]{Sims_2020}, and the null-detection from the SARAS~3 experiment~\citep{SARAS3}. 
Because of this lack of definitive detection, there are several ongoing and proposed experiments to try and measure the sky-averaged 21-cm signal, e.g.\ REACH~\citep{REACH}, PRIZM~\citep{PRIZM}, and EDGES 3.
The question then naturally arises, what is the \textit{a priori} expectation of a global signal experiment with given sensitivity making a definitive detection of the uncertain 21-cm signal? 
Since a significant signal detection can be determined to occur when $\br$ between a model with a signal and a model without a signal exceeds some threshold $\br_{\rm crit}$, this question can be rigorously answered through a fully Bayesian forecast of the form given in \cref{eqn:fully_bayesian_distinguish}.

Hence, following our fully Bayesian forecasting methodology outlined above, we began by constructing the two mock data simulators, one that modelled a global 21-cm signal and one without. 
We imagined a REACH-like global 21-cm signal experiment with frequency-band covering redshift $7.5$ to $28.0$ and an analysis spectral resolution of $1$\,MHz.
For both simulators, we used the physically motivated galactic and ionospheric foreground model from \citet{Hills_2018}, and assumed Gaussian white noise with magnitude $\sigma_{\rm noise} = 0.015$\,K\footnote{As the noise scales with spectral resolution $\Delta \nu$ as $1/\sqrt{\Delta \nu}$ this is equivalent to $0.047$\,K noise at a resolution of $0.1$\,MHz, which lies between the \textit{pessimistic} and \textit{expected} case projected for REACH~\citep{REACH}.}.
For our global 21-cm signal model, we used \textsc{globalemu}~\citep{GLOBALEMU}, an emulator of a more computationally costly semi-numerical simulation code~\citep[e.g.,][]{Visbal_2012, Fialkov_2014, Reis_2020}. 
This model has seven parameters, star formation efficiency $f_{*}$, minimum circular virial velocity $V_{\rm c}$, X-ray emission efficiency $f_{\rm X}$, optical depth to the cosmic microwave background $\tau$, exponent $\alpha$ and lower-cutoff $E_{\rm min}$ of the X-ray spectral energy distribution, and the maximum root-mean free path of ionizing photons $R_{\rm mfp}$.
We specified our \textit{a priori} knowledge of these parameters and the foreground parameters through our priors listed in \cref{tab:priors}. 
The astrophysical priors used are uniform or log-uniform distributions centred on theoretically expected values, except for $\tau$, which we used a truncated Gaussian prior based on the \textit{Planck} 2018 posterior on $\tau$~\citep{Planck_VI}.
For the foreground parameter priors, we used physically restricted priors following \citet{Hills_2018}.
Thus, by combining noise generators, our analytic foreground model, \textsc{globalemu}, and samplers over our priors, we constructed simulators of mock global 21-cm signal data from a model with only noise and foreground (no-signal) and a model with noise, foreground, and signal (with-signal).  
Our two competing models thus have moderate dimensionalities (5 and 12) and depend non-linearly on their parameters, leading to complex data manifolds. 
In addition, this particular classification task is made more challenging due to the large shared foreground component of the two models being $10^4$ to $10^5$ times larger than the signal.

\begin{table}
 \caption{Priors on our foreground and 21-cm global signal parameters. To keep the parameter values within the training parameter ranges of \textsc{globalemu} we use a truncated Gaussian prior on $\tau$ derived from the Planck 2018 measurements rather than a Gaussian prior.}
 \label{tab:priors}
 \begin{ruledtabular}
 \begin{tabular}{cccccc}
  Parameter & Prior Type & Min & Max & Mean & Std. \\
\colrule
  $d_{\rm 0}$ (K) &  Uniform  & 1500 & 2000 & - & - \\
  $d_{\rm 1}$ &  Uniform  & -1.0 & 1.0 & - & - \\
  $d_{\rm 2}$ &  Uniform  & -0.05 & 0.05 & - & - \\
  $\tau_{e}$  &  Uniform  & 0.005 & 0.200 & - & - \\
  $T_{e}$ (K)  &  Uniform  & 200 & 2000 & - & - \\
\colrule
  $f_{*}$ & Log Uniform & 0.0001 & 0.5 & - & - \\
  $V_{\rm c}$ (km\,s$^{-1}$) & Log Uniform & 4.2 & 30.0 & - & - \\
  $f_{\rm x}$ & Log Uniform & 0.001 & 1000.0 & - & - \\
  $\tau$ & Truncated Gaussian & 0.040 & 0.17 & 0.054 & 0.007 \\
  $\alpha$ & Uniform & 1.0 & 1.5 & - & - \\
  $E_{\rm min}$ (keV) & Log Uniform & 0.1 & 3.0 & - & - \\
  $R_{\rm mfp}$ (cMpc) & Uniform & 10.0 & 50.0 & - & - \\
 \end{tabular}
 \end{ruledtabular}
\end{table}

To train our evidence network, we generated 32,000,000 (12,800,000) mock data sets from each simulator to form our training (validation) set.  
Our evidence network was implemented in \textsc{tensorflow}~\citep{tensorflow}, and consisted of an initial Cholesky whitening transform~\citep{Kessy_2015} followed by dense hidden layers of size 256-256-64-64-64-64-64-64-1, with batch normalization and a ReLU activation functions on all layers except for the output node, and an additive skip connection between the third and sixth layers to ease training.
We used the $\alpha = 2$ l-POP exponential loss function recommended for evident networks by \citet{EN} and the Adam optimizer with an initial learning rate of $10^{-3}$, decay steps of $10^5$, and decay rate of 0.95.

The network was trained with early-stopping for 900 epochs using a batch size of 32,768.

To validate that the network had converged to an accurate prediction of $\br$, we performed a blind coverage test as outlined in \citet{EN}.
This test consists of using the network to predict the posterior model probabilities for a range of test data sets, and then binning the data sets by these probabilities. 
If the network has correctly converged, the model probability corresponding to each bin should equal the proportion of the data sets in the bin generated from said model, which we indeed find to be the case for our testing set composed of 1,000,000 mock data sets generated from each model.
Furthermore, as in this case, we can construct an explicit likelihood; we additionally validated our network and methodology by comparing whether a significant detection would be concluded based on the network $\br$ values and $\br$ computed using \textsc{polychord}~\citep{POLYCHORD_I, POLYCHORD_II}.
For a signal detection threshold of 3\,$\sigma$ (5\,$\sigma$)\footnote{These $\sigma$ thresholds are translated to $\br_{\rm crit}$ values via equation~\eqref{eqn:model_posterior_absolute}, and requiring that the posterior probability of the model with no signal is less than the corresponding $p$ values of $2.70\times10^{-3}$ or $5.73\times10^{-7}$.}, corresponding to $\log(\br_{\rm crit}) = 5.91$ ($\log(\br_{\rm crit}) = 14.4$), we found the two methods came to the same conclusion on whether a signal was detected for 96.6\% (95.1\%) of 1000 mock data sets from our noisy-signal model, showing good agreement.

Finally, we evaluated \cref{eqn:fully_bayesian_distinguish} using our evidence network and 1,000,000 sets of noisy-signal model mock data.
Ultimately, we found the expected chance of the experiment detecting the global 21-cm signal at a statistical significance of 3\,$\sigma$ (5\,$\sigma$) was 46.0\% (32.4\%).
This approximately $50$\% change of a 21-cm signal detection at $>3$\,$\sigma$ confidence, suggests the $0.015$\,K sensitivity at 1\,MHz resolution considered here is indicative of the minimum sensitivity global signal experiments such target.

\begin{figure}
\includegraphics{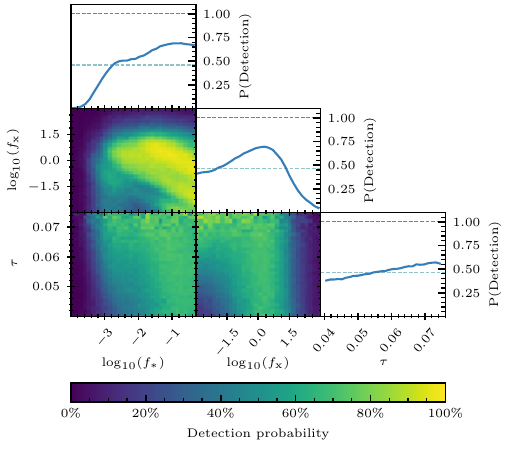}
    \caption{Triangle plot depicting the probability of a global 21-cm signal detection at $\geq 3$\,$\sigma$ statistical significance, by an experiment covering redshift $7.5$ to $28.0$, with frequency resolution $\Delta \nu = 1$\,MHz, assuming the \citet{Hills_2018} physical foreground model and white noise of \mbox{$\sigma_{\rm noise} = 0.015$\,K}.
The diagonal shows the total probability of a detection marginalized over noise realizations and the parameter space except for one fixed parameter, and the below diagonal is the equivalent with two fixed parameters. 
The total detection probability with no parameters fixed was $46.0$\%.
$V_{\rm c}$, $\alpha$, $E_{\rm min}$, and $R_{\rm mfp}$ are not shown because the probability of detection was found to vary only weakly with them. 
We find high $f_{\rm *}$, $f_{\rm X} \approx 1$, and high $\tau$ values increase the chance of a 21-cm signal detection, with $f_{\rm *}$ the most impactful.  
For different parameter combinations, we find detection probabilities varying from $\sim 0$\% to $\sim 100$\%, illustrating the need for fully Bayesian forecasts as the conclusion of detectability varies strongly over the \textit{a priori} uncertain parameter space. }
\label{fig:headline_plot}
\end{figure}

Additionally, our 1,000,000 $\br$ evaluations allow for a broad range of further analyses at little to no additional computation cost. 
Instead of performing the average over $\pi$ in \cref{eqn:fully_bayesian_distinguish}, we can marginalize over conditional priors with one or two parameters fixed, giving insight into under which early Universe astrophysical scenarios we would expect to detect the 21-cm signal.
These conditional detection probabilities are depicted in \cref{fig:headline_plot} for the parameters with which strong variation in detection probability was seen, $f_{*}$, $f_{\rm X}$, and $\tau$. 
We find the detection of the 21-cm signal is more likely for high star formation efficiencies, X-ray efficiencies around the theoretically expected value of 1, and higher optical depths to reionization. 
With an almost 100\% chance of a 3\,$\sigma$ detection for $f_{*} >0.01$ and $f_{\rm X} = 1$, and an effectively 0\% chance of  a 3\,$\sigma$ detection for $f_{*} < 0.001$ or $f_{\rm X} > 30$. 
This strong variation in definitive detection chances retroactively provides further motivation for fully Bayesian forecasts as we see the conclusion drawn would be highly sensitive to the fiducial global 21-cm signal parameters chosen for a traditional Bayesian forecast. 

There are a myriad of further analyses we could perform with our methodology. 
For example, we could study the evolution of the detection probability against the experiment noise level or bandwidth, which could, in turn, help inform experimental design.
Additionally, we could consider the functional distribution of the detected and non-detected 21-cm global signals to gain insight into what differentiates these two categories.
Or alternatively, we could investigate how the chances of a 21-cm global signal detection have changed in light of the parameter constraints from the HERA 21-cm power spectrum limits~\citep{HERA_theory_22, HERA_obs_23}, or the SARAS~3 null detection~\citep{SARAS3, Bevins_2022}.
Furthermore, if this were a problem where particular parameter(s) values were of special interest (e.g., the minimum sum of the neutrino masses permitted by particle physics) this methodology also allows for determining detection chance with the parameter(s) fixed to that value while still marginalizing over noise and nuisance parameters.
However, we shall leave such analyses to future work since the focus of this letter is on the method and its feasibility rather than particular scientific problems.

Let us return now to the computational performance of our method.
The training data generation, network training,  forecast data generation, network $\br$ evaluations, and plotting of \cref{fig:headline_plot}, took a combined total of 5.54\,GPU hours\footnote{On an NVIDIA A100-SXM-80GB GPU that was part of a \href{https://docs.hpc.cam.ac.uk/hpc/user-guide/a100.html}{CSD3 HPC Ampere GPU node}.}.
Conversely, the 1000 \textsc{polychord} evaluations of $\br$ as part of our validation process took a total of 45,000\,CPU hours\footnote{On 76 Intel Ice Lake CPUs that form a \href{https://docs.hpc.cam.ac.uk/hpc/user-guide/icelake.html}{CSD3 HPC Ice Lake node}.}, from which we can estimate the 1,000,000 $\br$ evaluations used in our fully Bayesian forecast would have required 45,000,000\,CPU hours using traditional methods. 
While it is not meaningful to directly compare GPU to CPU hours we can compare the costs of those hours.
On the cluster we utilized GPU hours are charged at 50 times the rate of CPU hours. 
Thus, our method gives a cost-weighted performance improvement of $10^{5.2}$. 
Since this performance gain was for a single problem, and our implementation was not optimized, this level of performance gain cannot be assumed to apply universally. 
However, it is indicative our methodology is highly performant compared to traditional techniques and, as we have directly demonstrated, facilitates analyses that were previously computationally prohibitively expensive.

\textit{Conclusions.}---
We have argued, like \citet{Mukherjee_2006} and \citet{Trotta_2007b} before us, that to arrive at accurate and interpretable predictions, the conclusions of scientific forecasts should be marginalized over any uncertain model parameters and noise realizations.
However, such fully Bayesian forecasts are computationally infeasible with traditional methods for model comparison forecasts. 
We thus propose a novel methodology for performing fully Bayesian forecasts based on Evidence Networks.

To illustrate our method and the insights that can be gained from fully Bayesian forecasts, we applied it to determine the chances of a REACH-like experiment detecting the global 21-cm signal from beneath foregrounds and noise.
For a frequency resolution of $1$\,MHz and a noise level of $\sigma_{\rm noise} = 0.015$\,K, we find a 46.0\% (32.4\%) chance of detection at 3\,$\sigma$ (5\,$\sigma$). 
Thus suggesting this noise level is indicative of the minimum sensitivity global 21-cm signal experiments should target. 
Additionally, our methodology allows us to produce triangle plots of how this chance of detection varies when one or two model parameters are fixed, at no extra computational cost. 
For this example problem, we find a cost-weighted speed-up of $10^{5.2}$ using our approach compared to a traditional nested-sampling-based method that would have taken $45,000,000$\,CPU hours.

The method we propose can be applied to any forecasting question which can be formulated as a condition on the Bayes ratio between two models. 
This includes: if a signal can be detected from within noise (e.g.\ gravitational waves~\citep{CosmicExplorer}, or the 21-cm signal~\citep{Furlanetto_2006}); whether two competing theories can be distinguished by anticipated data (e.g.\ MOND~\citep{MOND} or General Relativity~\citep{GR}); or if the inclusion of novel physics in a model will be necessary (e.g.\ neutrinos in CMB experiment analysis~\citep{Simons}). 
Additionally, the method only requires simulators of mock data, and thus can still be used in cases where closed-form likelihoods or explicit priors are not available. 
As a result, this methodology should allow for reliable and efficient fully Bayesian forecasts on a wide range of forecasting problems.

Since the proposed methodology is simulation-based and has a low computational cost, we anticipate it will be particularly suited to performing forecasts for a range of potential experimental configurations. 
Thus allowing for the optimization of experimental configurations to minimize cost or maximize the chance of the detection of new physics. 
To facilitate the application of this methodology by others, we make \href{https://github.com/ThomasGesseyJones/FullyBayesianForecastsExample}{public} on GitHub all codes and data used in the writing of this letter.

\begin{acknowledgments}
We would like to express our gratitude to Harry Bevins for several insightful conversations concerning this work. 
The authors would like to thank the Science and Technology Facilities Council (UK) for their support of TGJ through grant number ST/V506606/1, and the Royal Society for their support of WJH  through a Royal Society University Research Fellowship. 
This work was performed using the Cambridge Service for Data Driven Discovery (CSD3), part of which is operated by the University of Cambridge Research Computing on behalf of the STFC DiRAC HPC Facility (www.dirac.ac.uk). The DiRAC component of CSD3 was funded by BEIS capital funding via STFC capital grants ST/P002307/1 and ST/R002452/1 and STFC operations grant ST/R00689X/1. DiRAC is part of the National e-Infrastructure.
For the purpose of open access, the author has applied a Creative Commons Attribution (CC BY) licence to any Author Accepted Manuscript version arising from this submission.
\end{acknowledgments}

\bibliographystyle{apsrev4-2}
\bibliography{paper}

\end{document}